\documentclass[journal]{IEEEtran}
\usepackage{graphicx}
\graphicspath{ {./} }
\ifCLASSINFOpdf
\else
\fi
\usepackage{authblk}

\begin{document}
%
\title{Coverage Enhancement for Vehicles}
%
%
%

\author[1]{Stefan~Runeson}
\author[2]{Ali~Zaidi}
\author[3]{Ana~Cantarero}
\affil[1]{Ericsson}
\affil[2]{Ericsson}
\affil[3]{BMW Group}

\maketitle

\begin{abstract}
The Third Generation Partnership Project (3GPP) has standardized Coverage Enhancement (CE) for Internet-of-Things (IoT) to connect devices in challenging radio conditions with cellular networks. CE is based on the principle of prolonged transmission time that exploits the fact that many IoT applications have relaxed requirements on data rate and latency, and the coverage can be significantly boosted by repeating transmissions for such applications. However, CE consumes a lot of radio resources and should be implemented carefully for different applications. This paper presents an end-to-end concept for realizing dynamic use of CE for connected vehicles in a resource efficient way. The proposed framework has the potential of improving coverage by around 10dB for low data rate connected vehicle applications, based on the 3GPP Long Term Evolution (LTE) standard.

\end{abstract}


%
\IEEEpeerreviewmaketitle

\section{Introduction}
%
%
%
%

\IEEEPARstart{T}{he} automotive industry is going through a fast-paced evolution, with the connected vehicle as one of its focus. A connected vehicle refers to one which can connect over wireless networks to other devices and its surroundings. This may go from vehicle-to-vehicle (V2V) and vehicle-to-infrastructure (V2I) communications, which are designed to increase situation awareness and road safety, to vehicle-to-network (V2N) communication, which supports a wider area over a cellular network. This paper is focused on V2N services which can be benefited from CE and how could this feature be enabled on a vehicular modem. 

Cellular V2N communication enables a wide range of services in the vehicle, including Cooperative Intelligent Transport Systems (C-ITS), Advanced Driver Assistance Systems (ADAS), dynamic high-definition maps, vehicle-centric Original Equipment Manufacturer (OEM) telematics, fleet management, logistics services and infotainment. Many of these services require high data rates and low latency for their reliability to be guaranteed. In this case, CE is not an option for optimization, due to the high amount of required radio resources. There are others however, whose data rate requirements are low and on which coverage and network connectivity should always be guaranteed. Here, CE could play a major role for optimization.  

These low data rate services include, among others, basic vehicle status updates to the OEM backend servers, remote vehicle lock or unlock and sharing of vehicle diagnostics, which are of major importance for fleet management and OEM logistics. Specially with the evolution of mobility concepts, such as car sharing, the fleet manager must have knowledge on the location and status of each vehicle on its fleet. The vehicles on the other side, should be able to connect to the backend servers when needed, even in areas of poor coverage. Areas such as underground garages or rural zones are of special interest, since their network coverage may be worse in comparison with urban zones or garages above ground.

\section{Concept description}
\subsection{3GPP standardization of Coverage Enhancement}

Third Generation Partnership Project (3GPP) has standardized Coverage Enhancement (CE) techniques for Internet-of-Things (IoT) devices in Release 13 and further evolved it in Release 14 \cite{CellularIoT}. Coverage Enhancement is primarily achieved by repetition or retransmission techniques for applications that have relaxed requirements on data rate and latency \cite{CellularIoT}. 

3GPP has standardized two CE modes: i) CE mode A, supporting up to 32 subframe repetitions with estimated coverage gain of around 10 dB, ii) CE mode B, supporting up to 2048 subframe repetitions. The repetitions are costly in terms of spectrum use, which is a scarce resource, and therefore this paper focuses on CE mode A and low data rate connected vehicle applications. 

There is a wide range of LTE UE categories supporting different capabilities and peak data rates \cite{LTEAdvanced}. CE mode A is mandatory for LTE Category-M1 UE, which is a very low complexity UE suitable for very low cost modems targeting low data rate connections (below 1Mbps peak rates). The connected vehicles typically employ higher category LTE UEs (capable of supporting tens or hundreds of Mbps) due to the need for delivering high data rates. CE mode A is optional for high category LTE UEs and is beneficial only when the device is out of normal LTE coverage and is supporting one or multiple low data rate applications. 

\subsection{Usage of Coverage Enhancement}

One way to take advantage of the benefits of CE in a vehicle would be to install an additional modem of LTE Cat-M1. This modem would then only be used for services with low requirements on data rates and latency. However, the cost would be increased with dual modems and the administrative effort would rise with dual Subscriber Identity Modules (SIMs).

The preferred concept is rather to use a single modem of high LTE UE category and allow the vehicle to dynamically enable the use of CE mode A in this modem.

When the vehicle is outside normal coverage and only use services with low requirements on data rates and latency, it can enable the use of CE mode A in the modem of high LTE UE category.

\begin{figure}[h]
\includegraphics[width=1.0\linewidth]{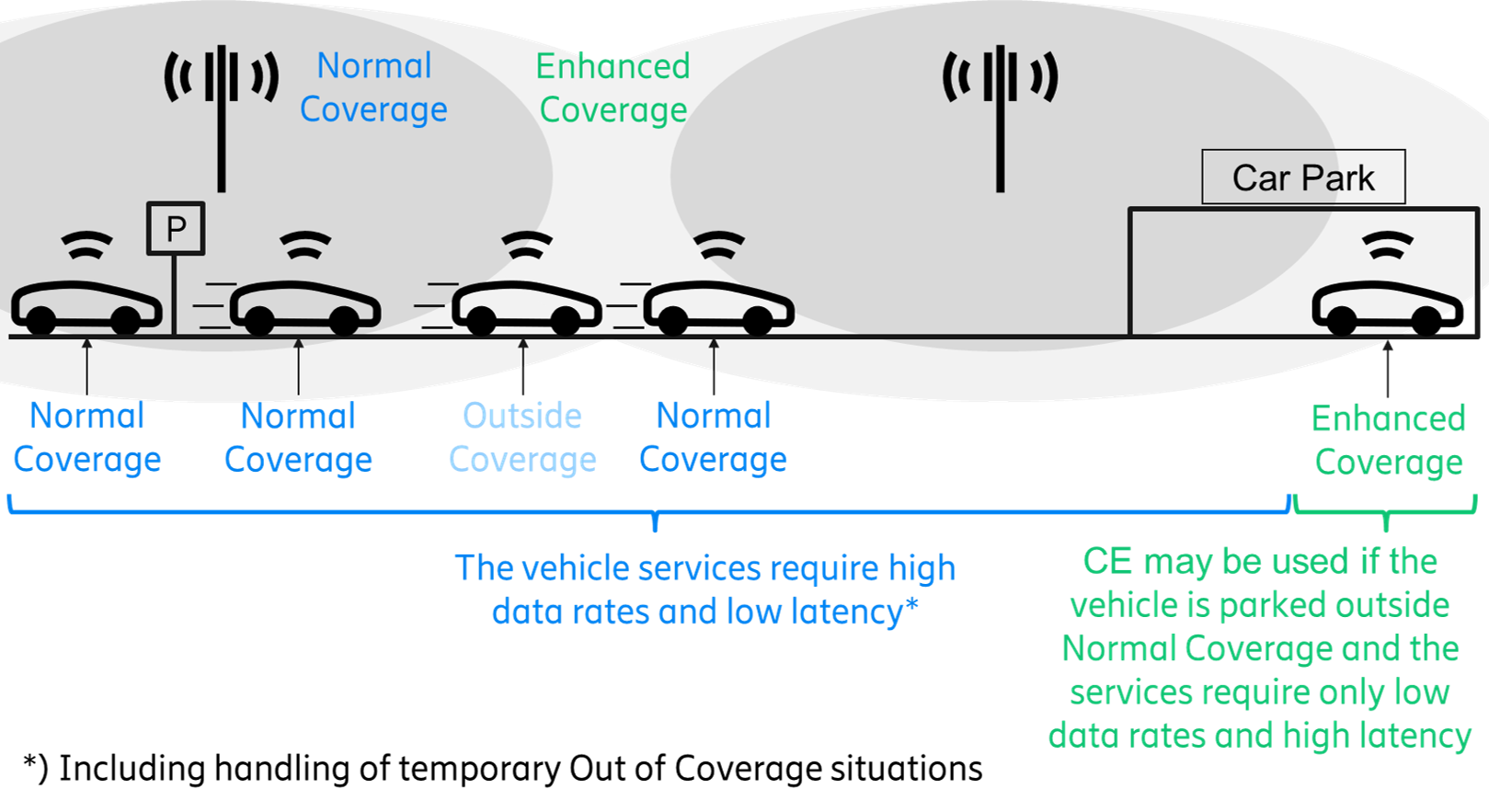}
\centering
\caption{Example of usage of Coverage Enhancement}
\label{fig:usage}
\end{figure}

One example where CE may be used is when the vehicle is parked outside normal coverage and only uses “remote services”\footnote{Transmission of a small information packet to the vehicle in order to trigger some action, such as starting the heating or flashing the headlights.} and “real time monitoring services”\footnote{Periodic transmissions from electrical vehicles to a backend server in order to track the current status of the battery.}, see Fig. \ref{fig:usage}.

Another example where CE may be used is an emergency situation outside normal coverage, where the vehicle may shut down other services and only send emergency messages.

\subsection{Enable and disable the use of CE mode A in the vehicle’s modem} 

Different UEs have different capabilities. The capabilities could be what frequency bands the UE supports, what radio features it supports, to name a few. The UE informs the mobile network of its capabilities when the UE attaches to the mobile network. The mobile network takes the individual UE’s capabilities into account when it interacts with the UE.

During the attach procedure, the base station (eNB) in the mobile network uses the Radio Resource Control (RRC) message “UE Capability Enquiry” to retrieve the UE capabilities from the UE. The UE capabilities are stored in the serving base station (eNB) but also in the Mobility Management Entity (MME), in order to keep the UE capabilities in the mobile network when the UE moves between different base stations.

\begin{figure}[h]
\includegraphics[width=1.0\linewidth]{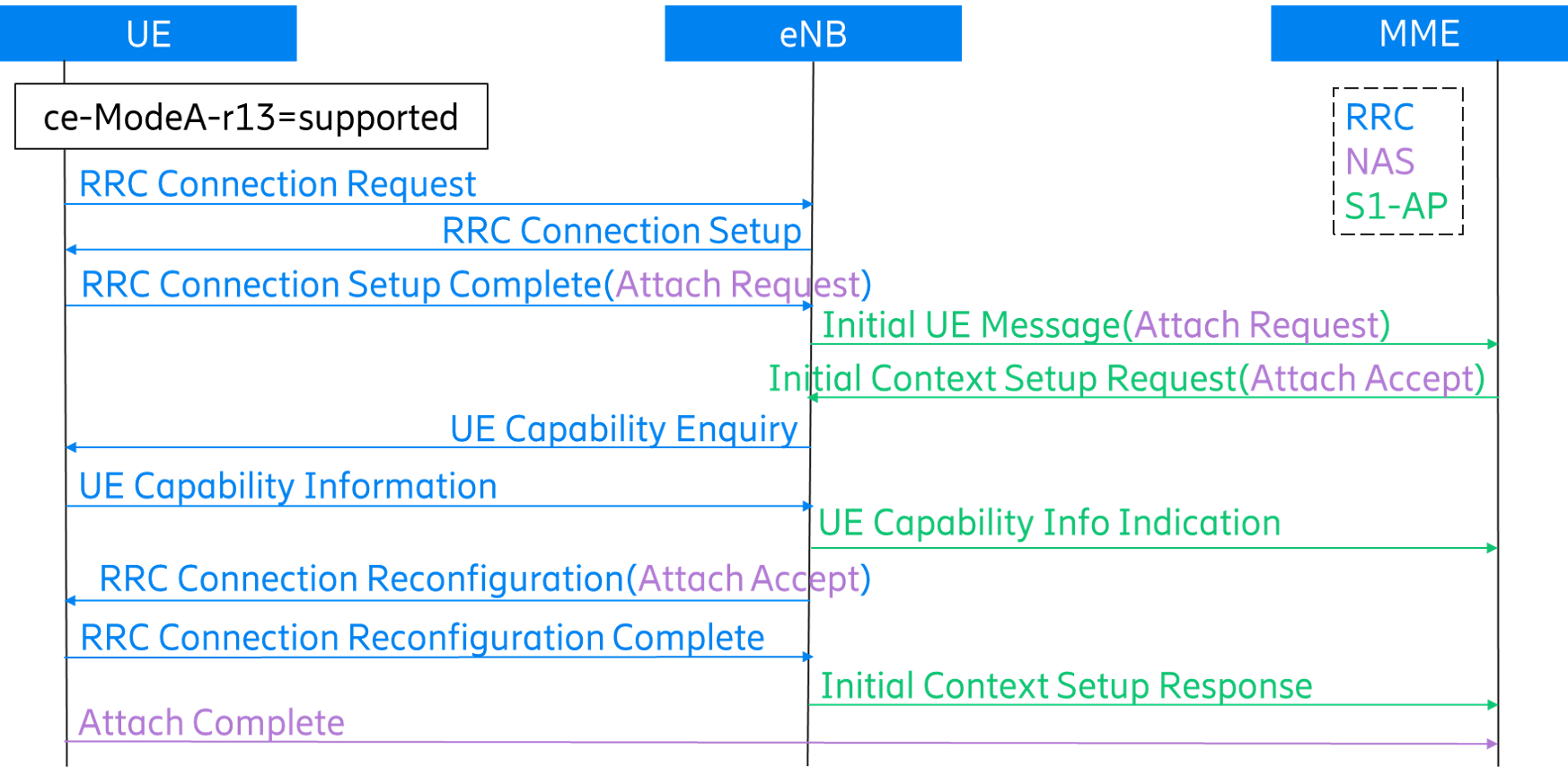}
\centering
\caption{3GPP procedures to enable support of CE mode A}
\label{fig:3GPPenable}
\end{figure}

Support for Coverage Enhancement is one such UE capability. The UE may indicate its support for CE mode A in its UE capabilities by setting the parameter “ce-ModeA-r13” to “supported” \cite{36.306}. In order to update the mobile network with new UE capabilities, the UE must start a new attach procedure, see Fig. \ref{fig:3GPPenable}. Radio Resource Control (RRC) is the control plane protocol between UE and eNB. Non-Access Stratum (NAS) is the control plane protocol between UE and MME. S1 Application Protocol (S1-AP) is the control plane protocol between eNB and MME.

\subsection{Restriction of use of Enhanced Coverage}

The usage of Enhanced Coverage may require use of extensive radio resources from the mobile network. 3GPP has standardized a restriction that enables the mobile network operator to prevent specific subscribers from using Enhanced Coverage \cite{23.401}.

The UE indicates its capability of support for restriction of use of Enhanced Coverage in the attach procedure to the MME. MME receives “Enhanced Coverage Restricted” parameter from the Home Subscriber Server (HSS). This parameter is retained as part of subscription data in the HSS and indicates whether the enhanced coverage functionality is restricted (i.e. disallowed) for the UE or not. 

For a UE that indicates support for restriction of use of Enhanced Coverage, MME sends a “Enhanced Coverage Restricted” parameter to the UE in the “Attach Accept” message. The UE assumes Enhanced Coverage is allowed unless explicitly restricted by the mobile network.

\begin{figure}[h]
\includegraphics[width=1.0\linewidth]{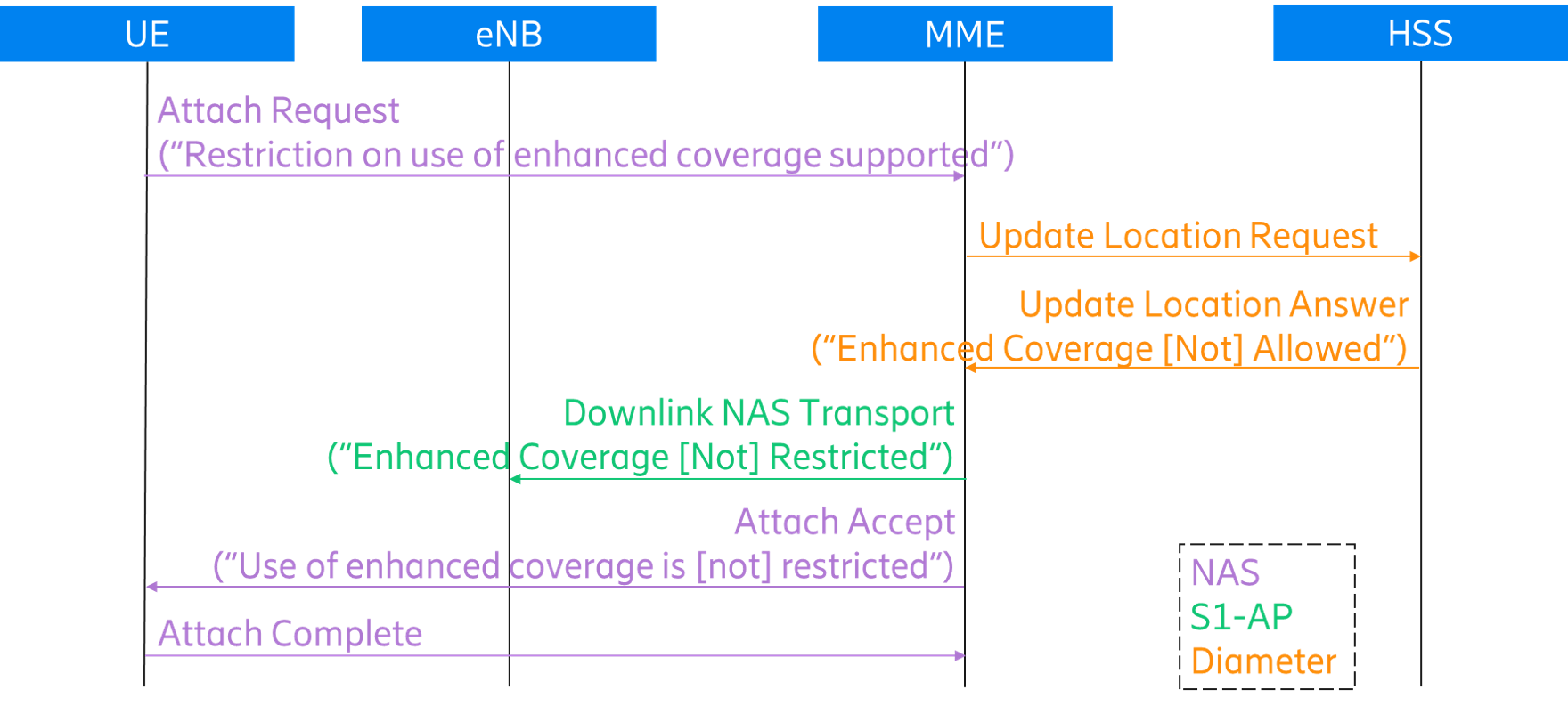}
\centering
\caption{3GPP procedures to restrict usage of Enhanced Coverage}
\label{fig:3GPPrestrict}
\end{figure}

If the MME has sent the “Enhanced Coverage Restricted” parameter to the UE, the MME provides the “Enhanced Coverage Restricted” parameter to the eNB via S1 Application Protocol (S1-AP), see Fig. \ref{fig:3GPPrestrict}. Diameter is the name of an authentication, authorization, and accounting protocol for computer networks. It is used between MME and HSS.

Since the “Enhanced Coverage Restricted” parameter is kept as part of subscription data in the HSS, the mobile network operator has a possibility to control and also charge for the Coverage Enhancement service for individual subscriptions.

\subsection{Coverage Enhancement for roaming vehicles}

Roaming is an important aspect of all automotive services. Vehicles may be manufactured in one country, sold in another country, and driven in a third country.

3GPP has standardized the roaming interface between Home Public Land Mobile Network (H-PLMN) and Visited Public Land Mobile Network (V-PLMN) for Coverage Enhancement. S6a is the roaming interface between MME in the visited network and HSS in the home network.

There are a few prerequisites for usage of Coverage Enhancement in the visited network:

\begin{enumerate}
  \item eNB and MME in the visited network must support CE mode A for UEs of high LTE UE category.
  \item MME in the visited network must get the subscription data “Enhanced Coverage [Not] Allowed” from HSS in the home network for roaming UEs.
\end{enumerate}

There might also be a need for an update of the commercial agreement between the home network operator and the visited network operator regarding the roaming UE’s use of Coverage Enhancement in the visited network.

\subsection{End-to-end concept for usage of Coverage Enhancement}

Coverage Enhancement may be used when the vehicle is outside normal coverage and only uses services with low requirements on data rates and latency. On the other hand, Coverage Enhancement shall not be used when the vehicle uses services with high requirements on data rates and latency.

\begin{figure}[h]
\includegraphics[width=1.0\linewidth]{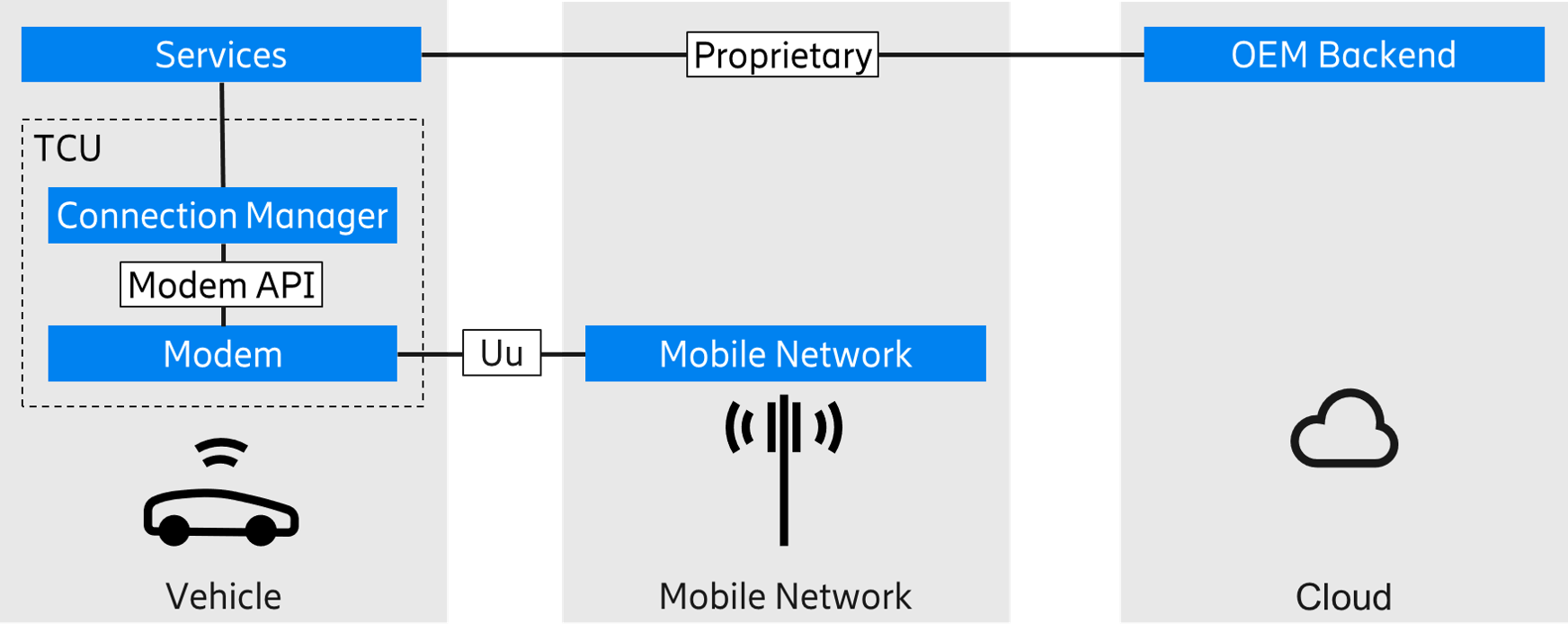}
\centering
\caption{End-to-end system using Coverage Enhancement}
\label{fig:E2Esystem}
\end{figure}

In order to use Coverage Enhancement in a proper way, an end-to-end concept for the system is needed. The system consists of three entities: vehicle, mobile network, and cloud. It is described in Fig. \ref{fig:E2Esystem}.

The services in the vehicle are connected via the mobile network to the OEM backend servers in the cloud. Proprietary application layer protocols are often used for these connections.

The vehicle has a Telematics Control Unit (TCU) with a modem and a connection manager.

The modem is connected to the mobile network via the 3GPP wireless interface (Uu). The modem establishes and monitors the connection to the mobile network. It exposes its services via an Application Programming Interface (API). The API could be based on AT-commands \cite{27.007}, or other types of interfaces.

The connection manager in the TCU establishes and monitors the connection from the vehicle to the OEM backend servers. In addition, the connection manager keeps track of whether the UE support CE mode A and whether any service in the vehicle needs data traffic with high requirements on data rates and latency.

\subsection{End-to-end procedure to enable the use of CE mode A}

\begin{figure}[h]
\includegraphics[width=1.0\linewidth]{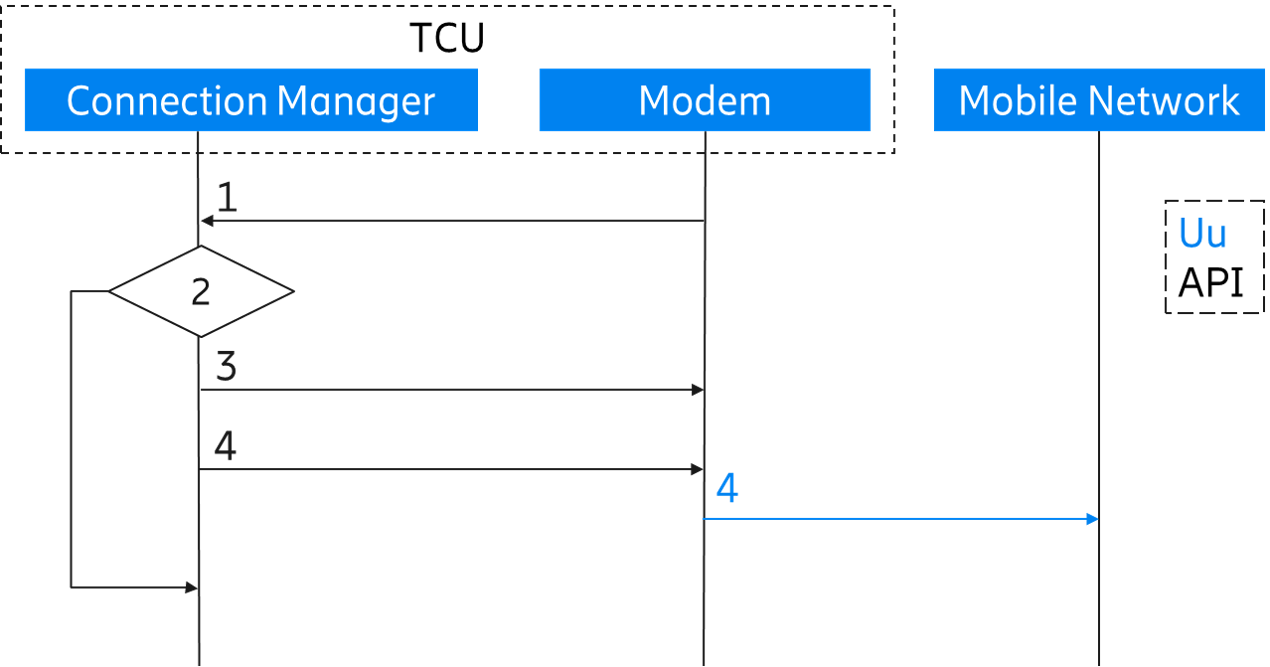}
\centering
\caption{End-to-end procedure to enable the use of CE mode A}
\label{fig:E2Eenable}
\end{figure}

The end-to-end procedure to enable the use of CE mode A consists of four steps, see Fig. \ref{fig:E2Eenable}.

\begin{enumerate}
  \item The connection manager detects that the vehicle is outside normal coverage. This could be based on information through the modem API.
  \item If one or more services in the vehicle need data traffic with high requirements on data rates and latency, it is not possible to use CE mode A and the procedure ends here.
  \item Otherwise, the connection manager updates the UE capabilities in the modem through the modem API to indicate that CE mode A is supported.
  \item The connection manager starts an attach procedure in the modem through modem API\footnote{If AT-commands are used as modem API, the “AT+CGATT=1” may be used \cite{27.007}.}.
\end{enumerate}

\subsection{End-to-end procedure to disable the use of CE mode A}

\begin{figure}[h]
\includegraphics[width=1.0\linewidth]{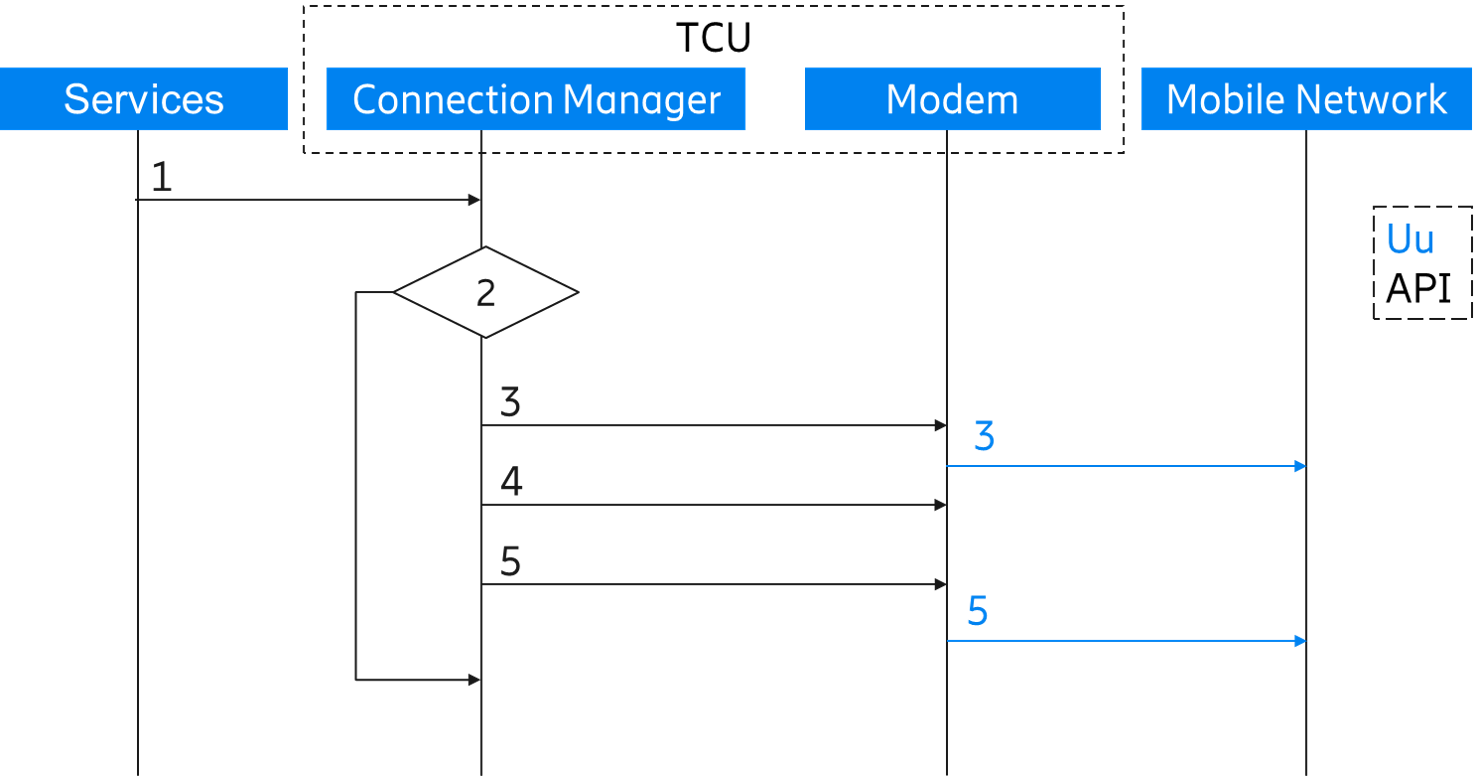}
\centering
\caption{End-to-end procedure to disable the use of CE mode A}
\label{fig:E2Edisable}
\end{figure}

The end-to-end procedure to disable the use of CE mode A consists of five steps, see Fig. \ref{fig:E2Edisable}.

\begin{enumerate}
  \item A service notifies the connection manager that the service needs data traffic with high requirements on data rates and latency.
  \item If the UE capabilities in the modem has no support for CE mode A, data traffic with high requirements on data rates and latency is permitted and the procedure ends here.
  \item Otherwise, the connection manager starts a detach procedure in the modem through modem API\footnote{If AT-commands are used as modem API, the “AT+CGATT=0” may be used \cite{27.007}}.
  \item The connection manager updates the UE capabilities in the modem through the modem API to indicate that CE mode A is not supported.
  \item The connection manager starts an attach procedure in the modem through modem API.
\end{enumerate}

\section{Conclusion}

The paper presented an end-to-end concept for enabling a dynamic use of CE for vehicles connected to cellular networks in a resource efficient manner. The proposed framework can be realized with LTE based on 3GPP Release 14 specification, with a potential of around 10 dB coverage improvement for low data rate applications. The framework presented in the paper can be generalized to other types of moving IoT devices, for example, drones and aerial vehicles. The concept can also be extended to 5G IoT devices, when the CE functionality is fully standardized for 5G in the future 3GPP releases.


%





\ifCLASSOPTIONcaptionsoff
  \newpage
\fi

\begin{IEEEbiographynophoto}{Stefan Runeson}
(stefan.runeson@ericsson.com) is a Senior Specialist Connected Vehicles at Ericsson Business Area Networks. He received his Master’s degree in Engineering Physics from Lund University, Sweden. He joined Ericsson in 1997 to work with data communications in mobile phones. Since 2016, he has been working with the impact of connected vehicles on mobile networks. Since 2017, he is Ericsson's delegate in the 5G Automotive Association (5GAA).
\end{IEEEbiographynophoto}

\begin{IEEEbiographynophoto}{Ali Zaidi}
(ali.zaidi@ericsson.com) is a strategic product manager for Cellular IoT at Ericsson and also serves as the company’s head of IoT Competence. Since joining Ericsson in 2014, he has been working with technology and business development of 4G and 5G radio access at Ericsson. He is currently responsible for LTE-M, URLLC, Industrial IoT, vehicle-to-everything and local industrial networks. He holds a Ph.D. in telecommunications from KTH Royal Institute of Technology, Stockholm.
\end{IEEEbiographynophoto}

\begin{IEEEbiographynophoto}{Ana Cantarero}
(ana.cantarero@bmw.de) received her Master’s degree in Communications Technology from the University of Ulm, Germany in 2012. In her position at BMW Group, her field of research is the connected vehicle with recent emphasis on Vehicle-to-everything (V2X) communication and future wireless communication technologies. She is currently involved in several national and European projects related to cellular communication networks contributing to the groundwork of 5G. Before joining BMW Group, she worked as Radio Access Network Planning Engineer at Millicom International Cellular and as Software Developer for High-Frequency 4G and 5G Test Systems at Rohde \& Schwarz. 
\end{IEEEbiographynophoto}






\end{document}